\documentclass[toc]{PoS}
\usepackage{wasysym}   

\title{Low energy kaon-nuclei interaction studies through the 
$\Sigma^0\pi^0$ channel with the KLOE detector}

\ShortTitle{Low energy kaon-nuclei interaction studies through the 
$\Sigma^0\pi^0$ channel with the KLOE detector}

\author{\speaker{K. Piscicchia$^a$$^b$,}\\
C. Curceanu$^b$, A. Scordo$^{b}$, I.Tucakovic$^b$, O.Vazquez Doce$^c$,\\
on behalf of AMADEUS collaboration\\
\newline
{$^a$}Museo Storico della Fisica e Centro Studi e Ricerche "Enrico Fermi", Roma, Italy\\
{$^b$}INFN, Laboratori Nazionali di Frascati, Frascati (Roma), Italy\\
{$^c$}Excellence Cluster Universe, Technische Universit\"at M\"unchen, Garching,Germany\\
E-mail: \email{kristian.piscicchia@lnf.infn.it}}

\abstract{The AMADEUS experiment aims to perform precision studies in the sector of low-energy kaon-nuclei interaction at the DA$\Phi$NE collider at LNF-INFN, implementing a dedicated setup in the central region of the KLOE detector.
As a first step towards the AMADEUS realization the existing KLOE data (runs from 2002 to 2005) were analysed using the detector itself as an active target. $K^-$ nuclear interactions in the gas filling the KLOE drift chamber (Helium 90$\%$ and Isobutane 10$\%$) and the drift chamber entrance wall (mainly Carbon) were explored. Starting point was the reconstruction of the $\Lambda(1116)$ trough its decay into a proton and a pion ($BR = 63.9 \pm 0.5\%$).
Taking advantage of the good performances of the KLOE calorimeter in detecting photons we then focused on the investigation of the $\Lambda(1405)$, through its decay into $\Sigma^0\pi^0$.
The details of the $\Sigma^0 \pi^0$ analysis and preliminary results are presented.
}

\FullConference{International Winter Meeting on Nuclear Physics,\\
		21-25 January 2013\\
		Bormio, Italy }

\begin{document}

\section{Introduction}

This work is dealing with the study of the low-energy interactions of the negatively charged kaons with light nuclei.
Such type of physics, extremely important for the understanding of the non-perturbative  QCD in the strangeness sector, has important consequences going from hadron and nuclear physics to astrophysics.
In particular we investigated the $\Sigma^0 \pi^0$ channel, generated by $K^-$ absorptions on bound protons in ${}^4He$ and ${}^{12}C$. $\Sigma^0 \pi^0$ is the privileged, but still poorly explored \cite{ref:zychor, ref:prak,ref:moriya}, channel to investigate the yet unclear structure of the $\Lambda(1405)$ resonance, since it is free from the dominant $\Sigma(1385)$ ($I=1$) background.

The $\Lambda(1405)$ is generally accepted to be a spin $1/2$, isospin $I=0$ and strangeness $S=-1$ negative parity baryon resonance ($J^P=1/2^-$) assigned to the lowest $L=1$ supermultiplet of the three-quark system, together with its spin-orbit partner, the  ($J^P=3/2^-$) $\Lambda(1520)$.
Such state only decays into $(\Sigma \pi)^0$ ($I=0$) through the strong interaction.
Despite the fact that the $\Lambda(1405)$  has been observed in few experiments  and is currently listed as a four-stars resonance in the table of the Particle Data Group (PDG) \cite{ref:PDG}, its nature still remains an open issue.
The three quark picture ($uds$) meets some difficulties to explain both the observed $\Lambda(1405)$ mass and the mass splitting with the $\Lambda(1520)$.
The low mass of the $\Lambda(1405)$ can be explained in a five quark picture, which, however, predicts more unobserved excited baryons.
In the meson-baryon picture the  $\Lambda(1405)$ is viewed as a $\overline{K}N$ quasi-bound $I=0$ state, embedded in the $\Sigma \pi$ continuum, emerging in coupled-channel meson-baryon scattering models \cite{ref:dali3}.
A complete review of the broad theoretical and experimental literature is not possible here (see however \cite{ref:Hyodo}). The $\Lambda(1405)$ production in $\overline{K}N$ reactions is of particular interest due to the prediction, in chiral unitary models \cite{ref:kaiser,ref:oset,ref:oller}, of two poles emerging in the scattering amplitude (with $S=-1$ and $I=0$) in the neighborhood of the  $\Lambda(1405)$ mass. One pole is located at higher energy with a narrow width and is mainly coupled to the $\overline{K}N$ channel, while a second lower mass and broader pole is dominantly coupled to the $\Sigma \pi$ channel \cite{ref:weise2012}, and both contribute to the final experimental invariant mass distribution \cite{ref:oller,ref:nacher}.
Since the resonance is always seen in the invariant mass spectrum of the $\Sigma \pi$ strong decay, the only chance to observe a higher mass component is employing the $\overline{K}N$ production mechanism. Moreover the $\Sigma^0 \pi^0$ channel, which is free from the $I=1$ contribution and from the isospin interference term, turns to be the cleanest. Both objectives can be accomplished by using the KLOE detector at DA$\Phi$NE.

The $K^-$ nuclear absorption \emph{at rest} on ${}^4He$ and ${}^{12}C$ was explored, through the $\Sigma^\pm \pi^\mp$ channels, in bubble chamber \cite{ref:riley} and emulsion experiments \cite{ref:rif1,ref:rif2}. In the present study the contribution of the \emph{in flight} K$^-$ nuclear absorption to the  $\Sigma^0 \pi^0$ invariant mass spectrum was for the first time evidenced. The capture of low momentum kaons from DA$\Phi$NE on bound protons in  ${}^4He$ and ${}^{12}C$ allows to access a higher invariant mass region, above the threshold imposed by the proton binding energy, otherwise inaccessible.


\section{DA$\Phi$NE and the KLOE detector}

DA$\Phi$NE \cite{ref:dafne} (Double Anular $\Phi$-factory for Nice Experiments) is a double ring $e^+ \, e^-$ collider, designed to work at the center of mass energy of the $\phi$ particle $m_\phi = (1019.456 \pm 0.020) MeV/c^2$.
The $\phi$ meson decay produces charged kaons (with BR($K^+ \, K^-$) = $48.9 \pm 0.5 \%$) characterized by low momentum ($\sim 127$ MeV/c) which is ideal either to stop them in the materials of the KLOE detector, or to explore the products of the low energy nuclear absorptions of $K^-$. 

The KLOE detector \cite{ref:kloe} is characterized by a $\sim 4\pi$ geometry and an acceptance of $\sim98\%$; it consists of a large cylindrical Drift Chamber (DC) and a fine sampling
lead-scintillating fibers calorimeter, all immersed in an axial magnetic field of 0.52 T provided by a superconducting solenoid.
The DC \cite{ref:kloedc} has an inner radius of 0.25m, an outer radius of 2m, a length of 3.3m
and is centered around the interaction point. The DC entrance wall composition is 750 $\mu m$ of Carbon fibre and 150 $\mu m$ of aluminium foil.

Dedicated GEANT MC simulations  of the KLOE apparatus were performed to estimate the percentages of K$^-$ absorptions in the materials of the DC entrance wall (the K$^-$ absorption physics treated by the GHEISHA package). Out of the total fraction of captured kaons, about 81$\%$ results to be absorbed in the Carbon fibre component and the residual 19$\%$ in the aluminium foil.
The KLOE DC is filled with a mixture of Helium and Isobutane (90$\%$ in volume $^4$He and 10$\%$ in volume $C_4H_{10}$). the ratios of K$^-$ interactions in Helium ($N_{KHe}$) Carbon ($N_{KC}$) and Hydrogen ($N_{KH}$), giving rise to a $\Sigma^0\pi^0$ in the final state were estimated by using the corresponding absorption cross sections and branching fractions available in literature \cite{ref:van-vel,ref:deloff,ref:garcia,ref:van-vel2}. The results are 
$
\frac{N_{KHe}}{N_{KC}}  =  1.6 \pm 0.2 \; , \;  \frac{N_{KHe}}{N_{KH}}  =  570 \pm 71,
$
the contribution of $K^-H$ interactions turns then to be negligibly small.
The chamber is characterized by excellent position and momentum resolutions. 
Tracks are reconstructed with a resolution in the transverse $R-\phi$ plane
$\sigma_{R\phi}\sim200\,\mu m$ and a resolution along the z-axis $\sigma_z\sim2\,mm$.
The transverse momentum resolution for low momentum tracks ($(50<p<300) MeV/c$)
is $\frac{\sigma_{p_T}}{p_T}\sim0.4\%$.
The KLOE calorimeter \cite{ref:kloeemc} is composed of a cylindrical barrel and two endcaps,
providing a solid angle coverage of 98\%.
The volume ratio (lead/fibers/glue=42:48:10) is optimized for
a high light yield and a high efficiency for photons in the range
(20-300)MeV/c. The position of the cluster along the fibers can be obtained with a resolution $\sigma_{\parallel} \sim 1.4 cm/\sqrt{E(GeV)}$. The resolution in the orthogonal direction is  $\sigma_{\perp} \sim 1.3 cm$. The energy and time resolutions for photon clusters are given by $\frac{\sigma_E}{E_\gamma}= \frac{0.057}{\sqrt{E_\gamma (GeV)}}$ and 
$\sigma_t= \frac{57 ps}{\sqrt{E_\gamma (GeV)}} \oplus 100 ps$.

\section{The events selection}

\subsection{$\Lambda(1116)$ identification}

The analysis was performed on a sample of
about 1 fb$^{-1}$ integrated luminosity from the 2004-2005 KLOE data taking campaign.
In the present analysis charged kaons are identified using two-body decay  ($K^\pm \rightarrow \mu^\pm \nu$ or $K^\pm \rightarrow \pi^\pm \pi^0$) and/or the $dE/dx$ (the kaon identification is made by means of the ionization in the Drift Chamber gas) TAG mechanisms.

The optimization of the analysis cuts and resolution studies, was based on MC simulations of the quasi-free (non-resonant)  $\Sigma^0\pi^0$ production process. Both  K$^-$ absorptions in Carbon and Helium from the different components of the KLOE setup, were simulated.  K$^-$ is considered to interact with a single bound proton, the residual nucleus, left in its ground state, behaving like a spectator. Final state interaction was neglected. For each nuclear target and KLOE material both K$^-$ interactions at rest and in flight were considered. The influence of the energy loss on the helix described by the charged particles trajectory (from $\Sigma^0$ decay) in crossing various materials of the KLOE setup is taken into account.

First step of the analysis consists in the identification of a $\Lambda(1116)$ through the reconstruction of the decay vertex $\Lambda \rightarrow p + \pi^-$ (BR = 63.9 $\pm 0.5 \%$) which represents, for the $\Sigma^0\pi^0$ channel under consideration,  the signature of $K^-$ hadronic interaction.

\begin{figure}[h]
\centering
\vspace{-0.5cm}
\begin{tabular}{rl}
\mbox{\includegraphics*[height=7cm,width=7cm]{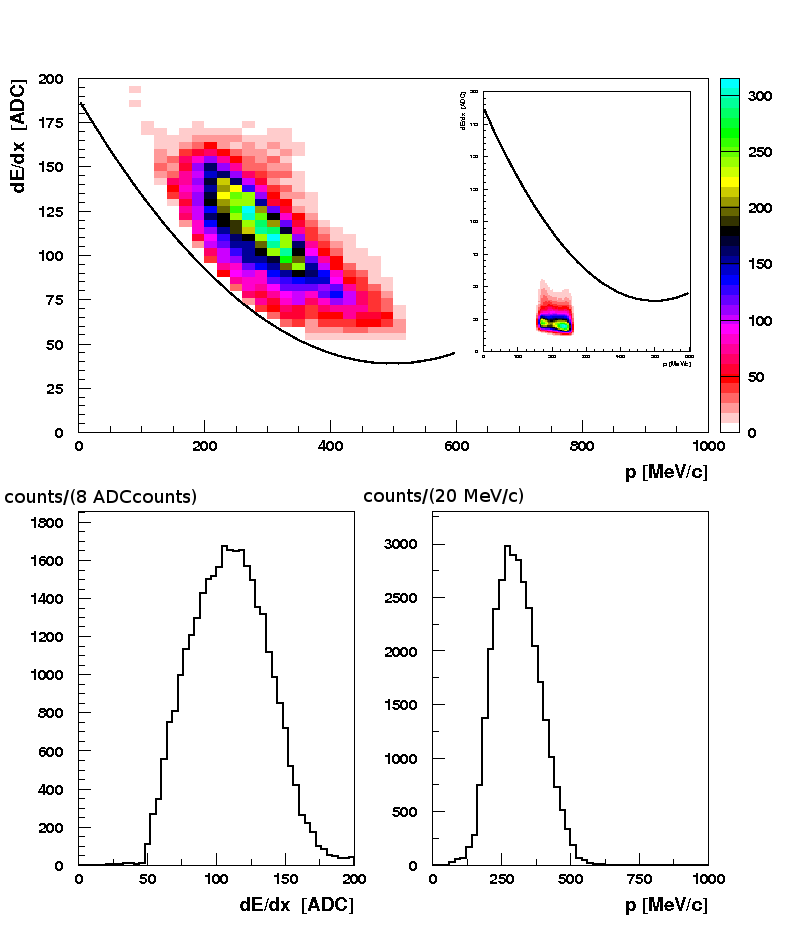}}
\mbox{\includegraphics*[height=7cm,width=7cm]{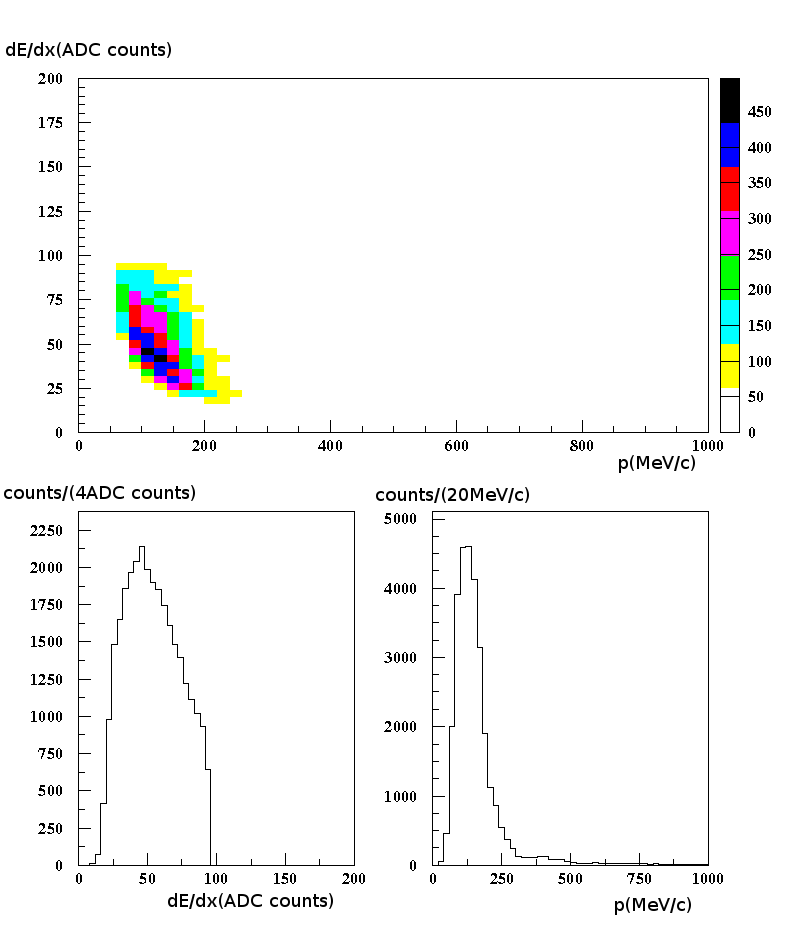}}
\end{tabular}
\caption{\em Left top: $dE/dx$ vs. momentum of the protons for the final selection (protons with + protons without associated cluster) (In the top right part, pions from 2-body decay are shown for comparison).
Left bottom: shows the corresponding projections (left: $dE/dx$, right: momentum).
Right: $dE/dx$ vs. momentum and
corresponding projections for the selected negative pions.
}
\label{finalp}
\end{figure}

In order to reduce the copious background from the three-body K$^{\pm}$ decays   ($K^\pm \rightarrow \pi^{\pm} \pi^{\pm} \pi^{\mp}$) we had
to refine the search criteria for the proton.
To this aim a cut on $dE/dx$ was optimized by characterizing
proton tracks with an associated cluster
in the calorimeter.
The signature of a proton in the calorimeter is clean,
since the corresponding signal is well-separated by the signal generated by the pions.
We were tuning the cut on these protons and then
applied the cut to all protons, i.e. including those which have the last DC measurement near the calorimeter (reaching it) but have no associated cluster. This last requirement enables to include in the selection low momentum protons (lower than $p \sim 250$ MeV/c) not producing an observable signal in the calorimeter. 
The cut is shown in Fig. \ref{finalp} left,
in which the momentum and $dE/dx$  of the finally selected protons
are plotted.
The function shown in the bidimensional plot shows the selection criterion.
This cut was optimized in order to reject positive pions contamination.
Inserted in the top right part of the same picture
we see the typical signature of pions in $dE/dx$ and momentum.
In this case we plotted $\pi^+$s
coming from the 2-body decay $K^+ \rightarrow \pi^+ \pi^0$.
This illustrates how the chosen cut is efficient for rejecting
$\pi^+$ in a broad range of momentum.

In Fig. \ref{finalp} right also the characteristics of the negative particle (pions)
of the vertices included in the final selection are shown.

As a final step for the identification of true $\Lambda$ decays the
vertices are
cross checked with quality cuts, using
the minimum distance between tracks and the chi-square
of the vertex fit.

\begin{figure}[h]
\centering
\vspace{-1.cm}
\begin{tabular}{rl}
\mbox{\includegraphics*[height=6.6cm,width=7.6cm]{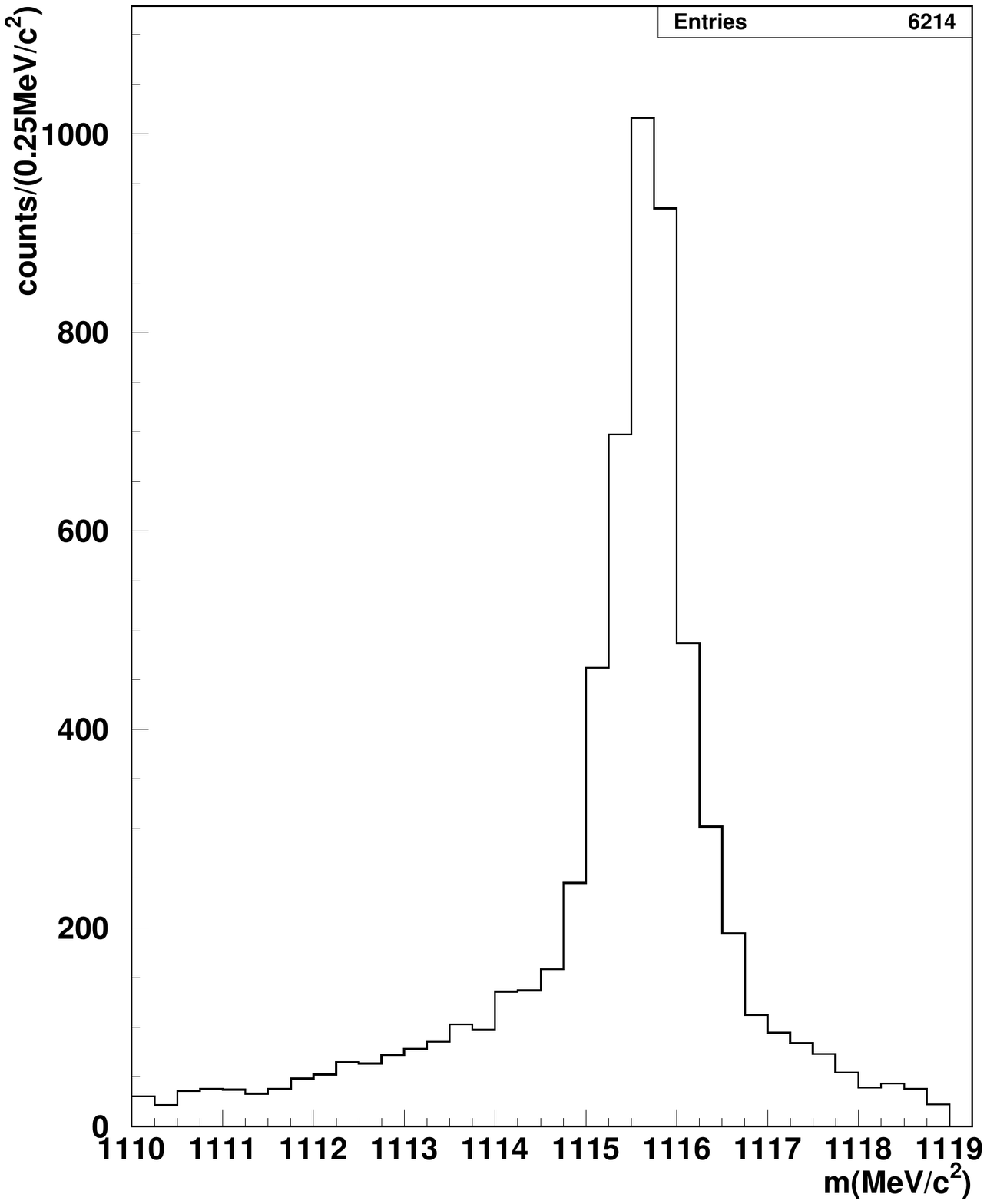}}
\mbox{\includegraphics*[height=7cm,width=7cm]{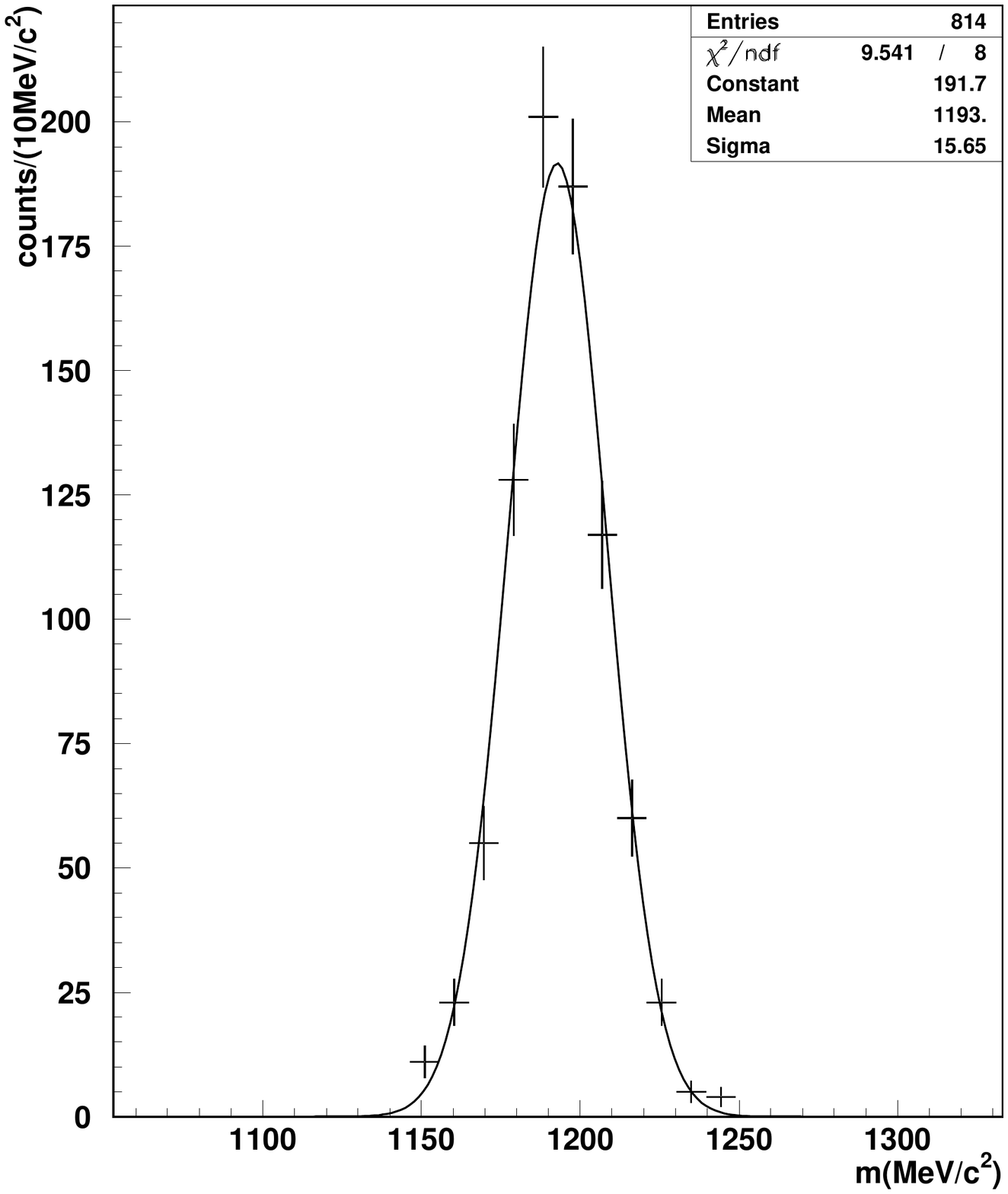}}
\end{tabular}
\vspace{-1.cm}
\caption{\em Left: $m_{p\pi^-}$ invariant mass spectrum. Right:  $m_{\Lambda\gamma 3}$ invariant mass distribution, together with a Gaussian fit.
}
\label{s1}
\end{figure}

The invariant mass $m_{p\pi^-}$ is calculated under the $p$ and  $\pi^-$ mass
hypothesis and is represented (for a preselected subsample of the total statistics) 
in Fig. \ref{s1} left. A cut is applied on the invariant mass ($1114 < m_{p \pi^-} < 1117$) $MeV/c^2$. Even if the introduced background of accidental pairs
lying under the lambda invariant mass is still considerable,
we keep these events in the analysis chain
since the additional requirements that will be introduced and
explained in the next section (namely the presence of a photon and a $\pi^0$ in time from the $\Lambda$ decay vertex) eliminate the background events almost
completely.

Cuts on the transverse radius ($\rho_\Lambda$) of the $\Lambda$ decay vertex were optimized in order to separate two samples of $K^-$ absorption events occurring in the DC wall and the DC gas: $\rho_\Lambda = 25 \pm 1.2 \, cm$ and $\rho_\Lambda > 30 cm$ respectively. The $\rho_\Lambda$ limits were set based on MC simulations and a study of the $\Lambda$ decay path. In particular the $\rho_\Lambda = 25 \pm 1.2 \, cm$ cut guarantees, for the first sample, a \emph{contamination} of $K^-$ interactions in gas as low as $(5.5^{+1.3}_{-1.8} \%)$, the dominance of $K^- \; {}^{12}C$ absorptions in the DC wall is fundamental for the comparative analysis performed with old emulsion experiments (see Section \ref{Comparing}). 

\subsection{$\Sigma^0 \pi^0$ identification}

The selection of $\Sigma^0 \pi^0$ events proceeds, after the $\Lambda(1116)$ identification, with the selection of three photon clusters. We will indicate as $\gamma_3$ the photon coming  from the $\Sigma^0$ decay, $\gamma_1$ and $\gamma_2$ will represent the photons from $\pi^0$ decay according to the reaction

\begin{equation}
K^- p \rightarrow \Sigma^0 \pi^0 \rightarrow (\Lambda \gamma_3) \, (\gamma_1 \gamma_2) \rightarrow (p\pi^-) \gamma_1 \gamma_2 \gamma_3
\end{equation}
A pseudo-chisquare minimization is performed, searching for three neutral clusters in the calorimeter ($E_{cl}>20MeV$), in time from the decay vertex position of the $\Lambda(1116)$ ($\textbf{r}_\Lambda$) ($\chi_t^2=(t_i-t_j)^2/\sigma_t^2$ where $t_i$ is the i-\emph{th} cluster time subtracted by the time of flight in the speed of light hypothesis).
According to dedicated MC simulations a cut was optimized on this variable $\chi_t^2 \le 20$.

Once the three candidate photon clusters are chosen, their assignment to the correct triple of photons ($\gamma_1,\gamma_2,\gamma_3$) is based on a second pseudo-chisquare minimization  ($\chi_{\pi\Sigma} ^2$). $\chi_{\pi\Sigma} ^2$ involves both the $\pi^0$ and $\Sigma^0$ masses. $\chi^2_{\pi\Sigma}$ is calculated for each possible combination and the minimizing triple is selected. The cut $\chi^2_{\pi\Sigma} \le 45$ was optimized based on MC simulations.

According to true MC information the algorithm has an efficiency of $(98\pm1) \%$ in recognizing photon clusters and an efficiency of $(78\pm1)\%$ in distinguishing the correct $\gamma_1 \gamma_2$ pair ($\pi^0$ decay) from $\gamma_3$ (the small misidentification source is to be ascribed to the calorimeter energy resolution and an overlapping of the photons energy  in a wide energy range).

A check is performed on the clusters energy and distance to avoid the selection of splitted clusters  (single clusters in the calorimeter erroneously recognized as two clusters) for $\pi^0$s. Cluster splitting is found to not affect significantly the sample.

In Fig. \ref{s1} right is shown the obtained invariant mass $m_{\Lambda\gamma 3}$ (for absorptions in the gas) together with a Gaussian fit. The resolution in the $m_{\Lambda \gamma 3}$ invariant mass is  $\sigma_{m_{\Lambda \gamma 3}} \sim 15 MeV/c^2$.
The resolutions on $\rho_\Lambda$ for the final selected $\Lambda$s are $\sigma_{\rho_\Lambda} \sim 0.20 cm$ (DC wall) and $\sigma_{\rho_\Lambda} \sim 0.13 cm$ (DC gas), corresponding to Gaussian fits to the distributions of the originally generated true-MC quantities subtracted by the reconstructed ones. The resolutions on $\Lambda$ momentum are  $\sigma_{p_{\Lambda}} \sim 4.5  MeV/c$ (DC wall) and $\sigma_{p_{\Lambda}} \sim 1.9 MeV/c$ (DC gas). The better resolution for the measured variables corresponding to K$^-$ hadronic interactions in the gas filling the KLOE DC is a consequence of the charged particles energy loss, mainly in the material of the DC entrance wall, particularly important for protons.  Such effect is completely overwhelmed in $\sigma_{m_{\Sigma^0\pi^0}}$  and $\sigma_{p_{\Sigma^0\pi^0}}$ by the calorimeter resolution for neutral clusters, each photon cluster introduces an enlargement of the order $10 MeV/c^2$ in the $\Sigma^0\pi^0$ invariant mass calculation (see section \ref{inv mass}).

\section{$\Sigma^0\pi^0$ invariant mass and momentum distributions}\label{inv mass}

\begin{figure}[h]
\centering
\vspace{-1.cm}
\begin{tabular}{rl}
\mbox{\includegraphics*[height=7cm,width=7cm]{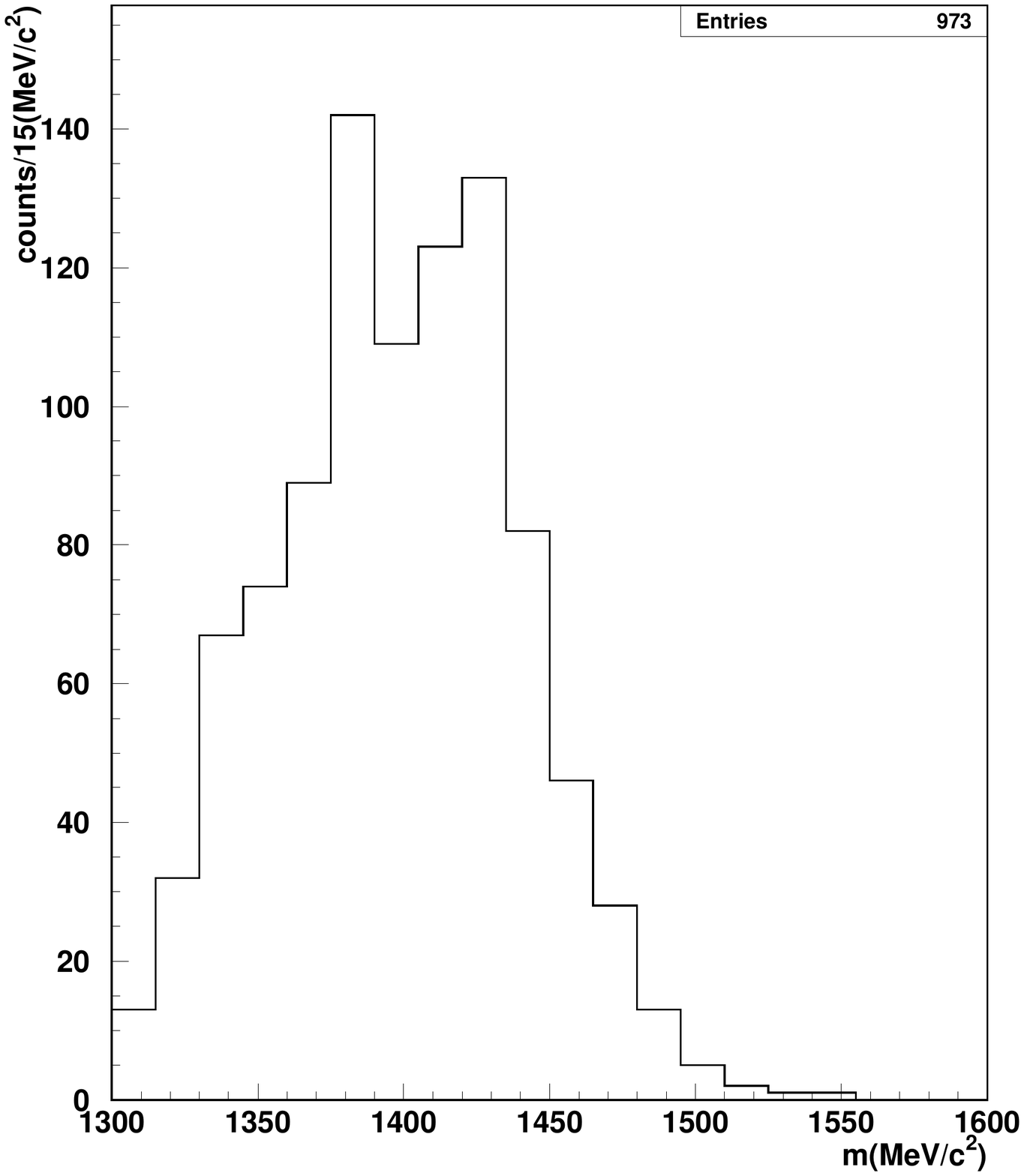}}
\mbox{\includegraphics*[height=7cm,width=7cm]{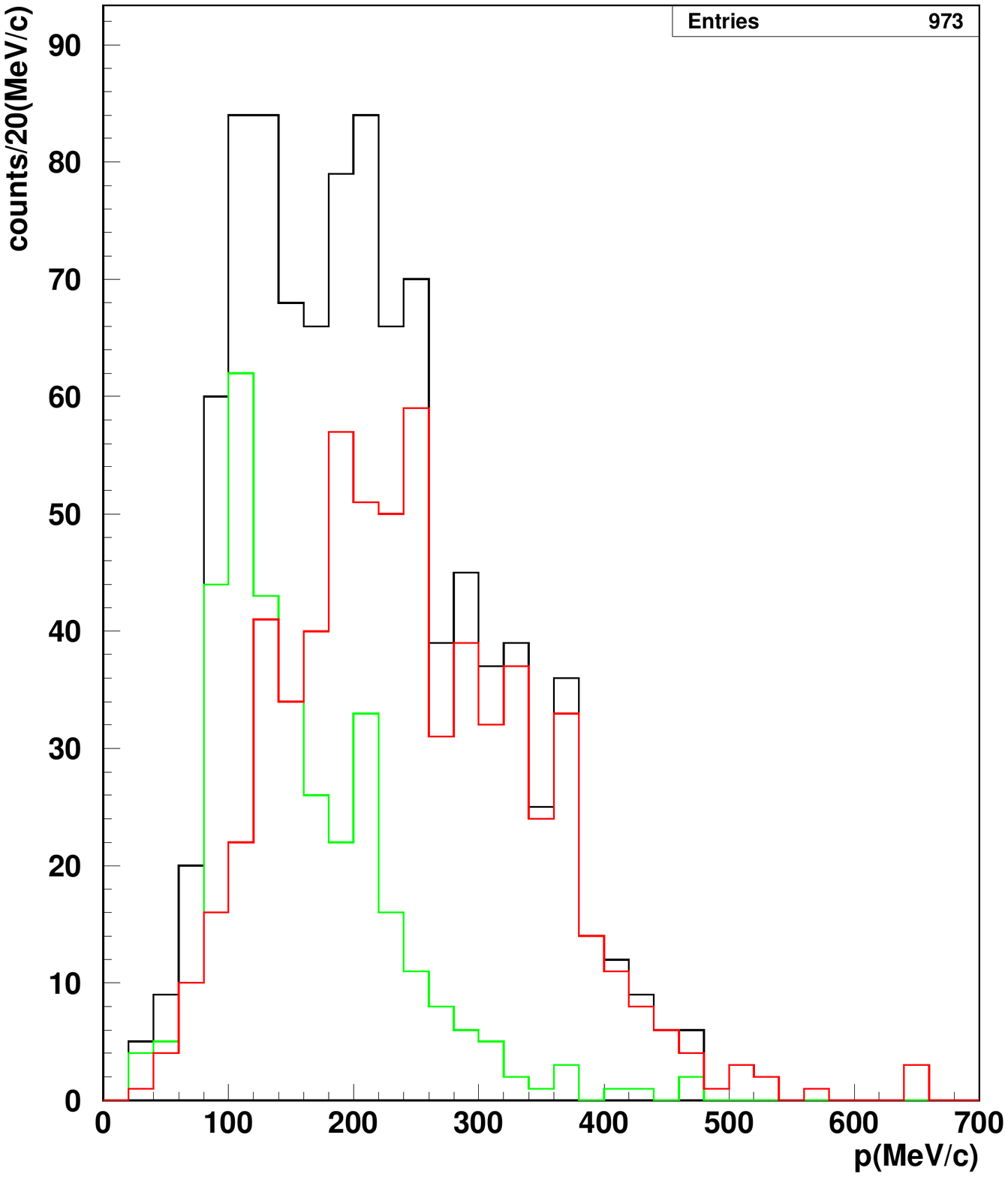}}
\end{tabular}
\vspace{-1.cm}
\caption{\em $m_{\Sigma^0\pi^0}$ left and $p_{\Sigma^0\pi^0}$ right for $K^-$ absorptions in the DC wall.
The red and green curves are obtained selecting events with $m_{\Sigma^0\pi^0}$ less or greater than $m_{lim}$ respectively.
}
\label{m_s0pi0_wall}
\end{figure}

\begin{figure}[h]
\centering
\vspace{-1.5cm}
\begin{tabular}{rl}
\mbox{\includegraphics*[height=7cm,width=7cm]{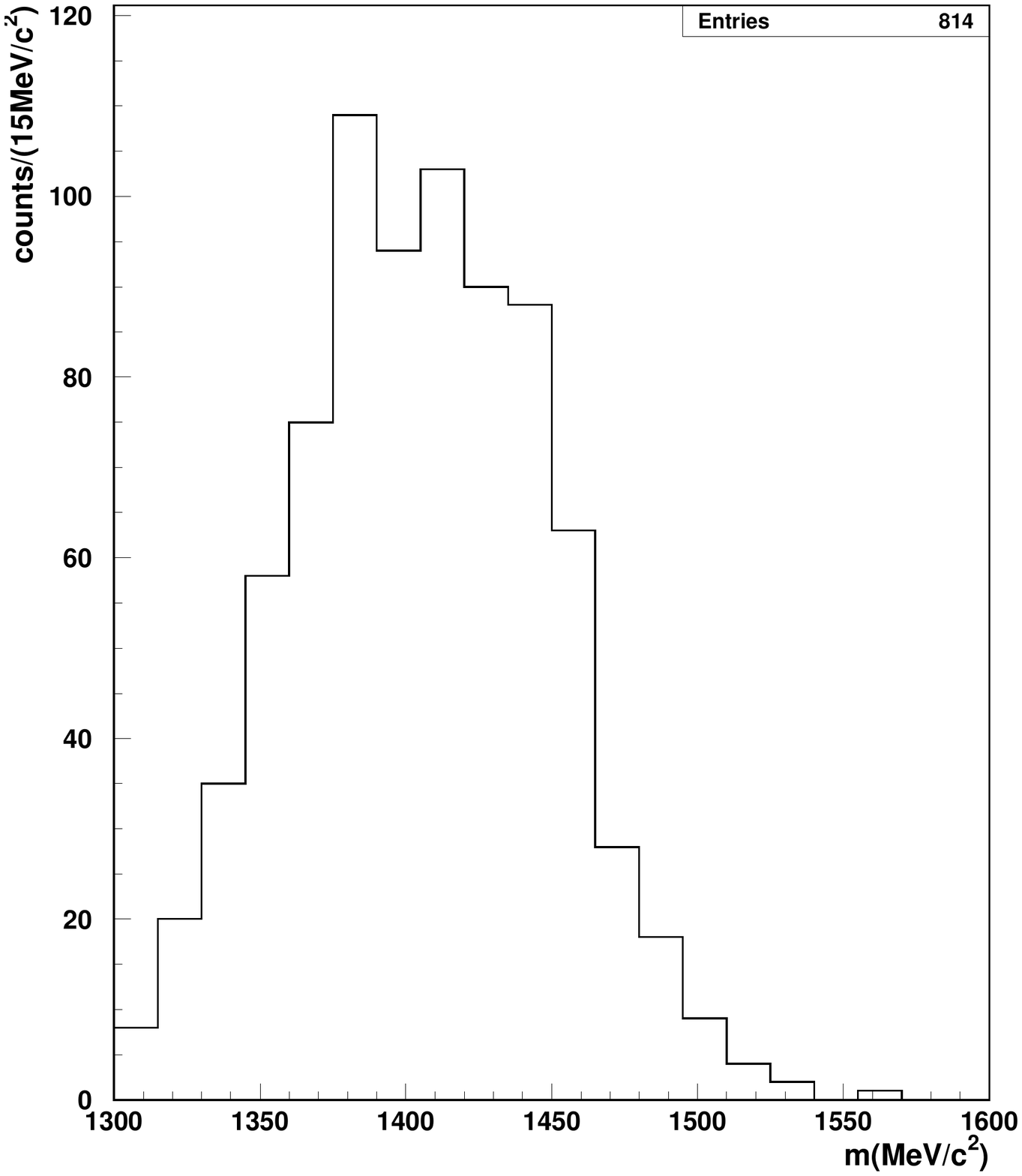}}
\mbox{\includegraphics*[height=7cm,width=7cm]{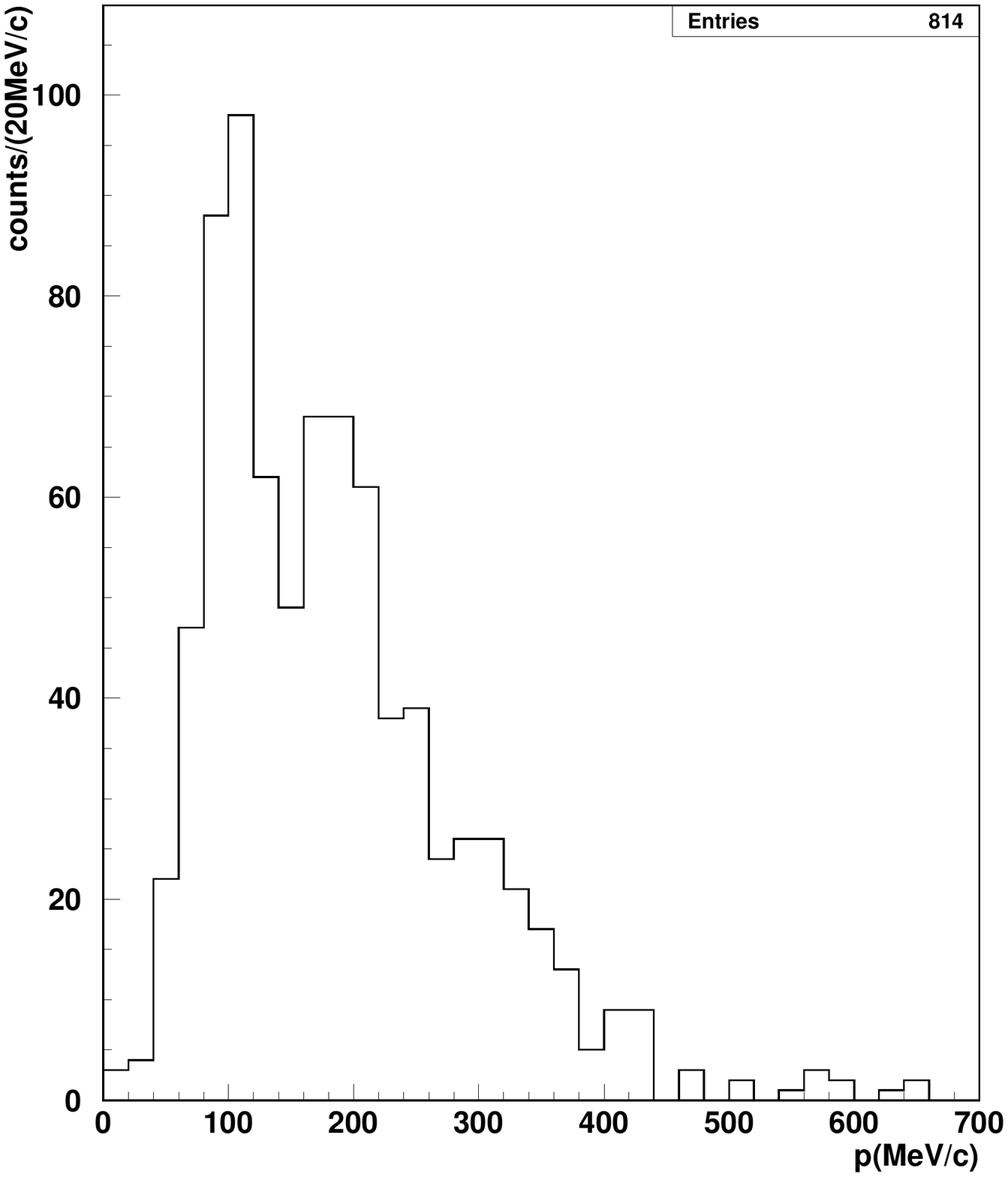}}
\end{tabular}
\vspace{-1.cm}
\caption{\em $m_{\Sigma^0\pi^0}$ left and $p_{\Sigma^0\pi^0}$ right for $K^-$ absorptions in the DC gas.
}
\label{m_s0pi0_gas}
\end{figure}

The $m_{\Sigma^0\pi^0}$ and $p_{\Sigma^0\pi^0}$ distributions (without mass hypothesis on the identified $\Sigma^0$s and $\pi^0$s) are represented in Figs \ref{m_s0pi0_wall} and \ref{m_s0pi0_gas} (black curves), for $K^-$ absorptions in the DC entrance wall and gas respectively. The corresponding resolutions are: $\sigma_{m_{\Sigma^0\pi^0}} \sim 32 MeV/c^2$ and $\sigma_{p_{\Sigma^0\pi^0}} \sim 20 MeV/c$ for $K^-$ absorptions in the DC wall, $\sigma_{m_{\Sigma^0\pi^0}} \sim 31 MeV/c^2$ and $\sigma_{p_{\Sigma^0\pi^0}} \sim 15 MeV/c$
for $K^-$ absorptions in the DC gas.

The $m_{\Sigma^0\pi^0}$ distributions are broad, and  show an excess of events above the kinematical limit for K$^-$ absorptions at rest (compared for example with the analogous $m_{\Sigma^\pm\pi^\mp}$ spectra in \cite{ref:riley,ref:rif2} which are sharply cut at the mass threshold). The kinematical limit $m_{lim}$ is almost the same for absorptions at rest in Carbon and Helium nuclei  ($\sim 1416MeV/c^2$  in Carbon, $\sim 1412MeV/c^2$ in Helium) as it depends on the last nucleon binding energy which differs only of 4 $MeV$ for the two nuclei. The $p_{\Sigma^0\pi^0}$ distribution shows a structure with two evident constituents, a lower momentum component (LM) at around $100 MeV/c$ and a higher momentum component (HM) around $200 MeV/c$.  The $p_{\Sigma^0\pi^0}$ spectrum is narrower for the events in the gas, probably as a consequence of the narrower protons Fermi distribution $p_{F}$ in  $^4$He (which accounts for most of the absorptions) respect to Carbon, which is reflected in $p_{\Sigma^0\pi^0}$. The small excess of events along the  high momentum tails for ($p_{\Sigma^0\pi^0}>250 MeV/c$) is probably associated to internal conversion events. More details about the study of this background source will be given in section \ref{back1}. 

\begin{figure}[h]
\centering
\vspace{-1.cm}
\begin{tabular}{rl}
\mbox{\includegraphics*[height=7cm,width=7cm]{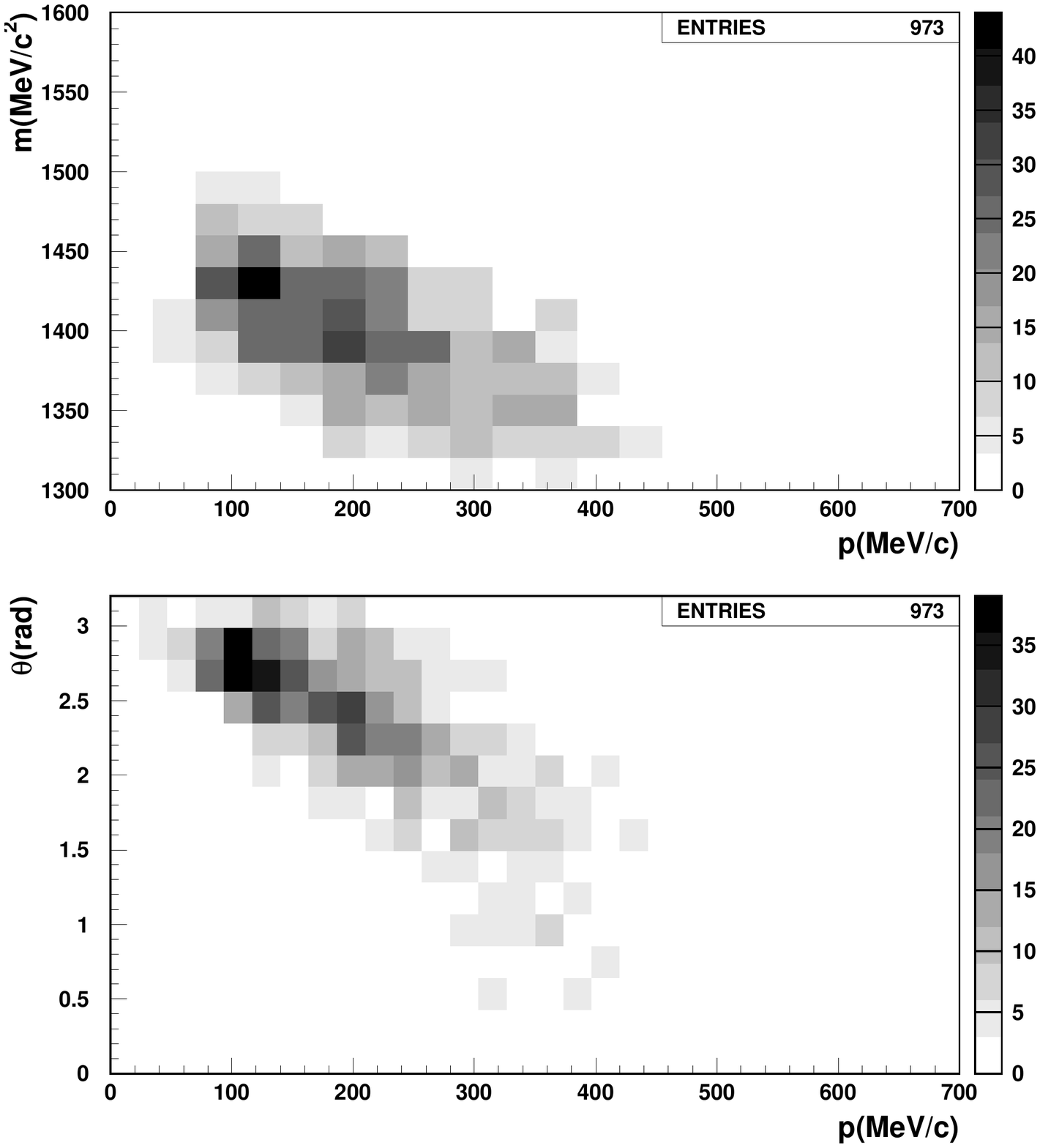}}
\mbox{\includegraphics*[height=7cm,width=7cm]{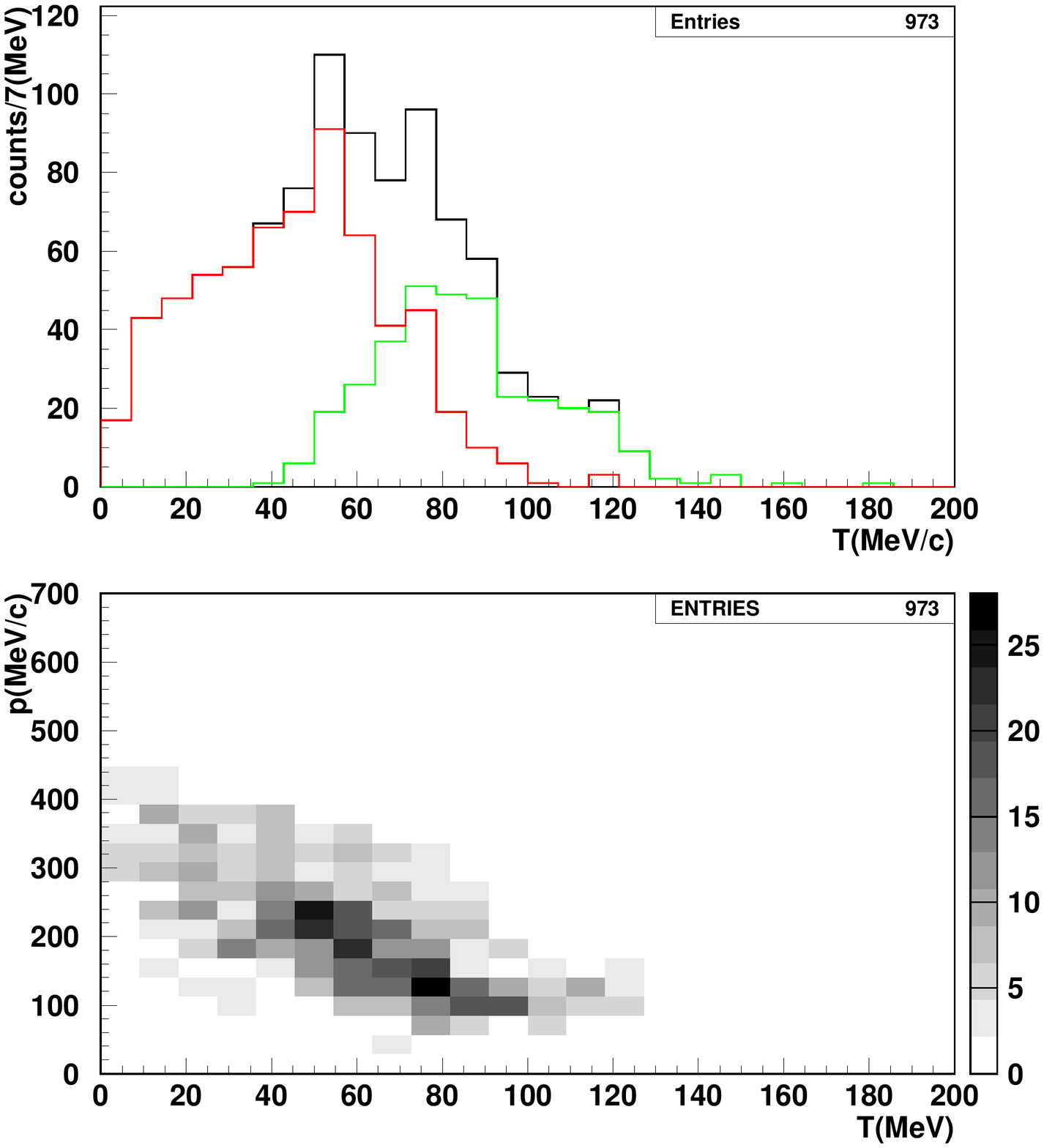}}
\end{tabular}
\vspace{-1.cm}
\caption{\em In the left panel are represented $m_{\Sigma^0\pi^0}$ vs. $p_{\Sigma^0\pi^0}$ (top), $\theta_{\Sigma^0\pi^0}$ vs. $p_{\Sigma^0\pi^0}$ (bottom). In the right panel is represented $p_{\Sigma^0\pi^0}$ vs. $T_{\pi^0}$ (bottom), together with the projection on the $p_{\Sigma^0\pi^0}$ axis (top).
The red and green curves are obtained selecting events with $m_{\Sigma^0\pi^0}$ less or greater than $m_{lim}$ respectively. 
}
\label{mtvsp_s0pi0}
\end{figure}

The LM and HM components are also rather clearly evidenced in the scatterplots of the relevant kinematical variables. In Fig.  \ref{mtvsp_s0pi0} left are shown the correlations of  $m_{\Sigma^0\pi^0}$ (top) and $\theta_{\Sigma^0\pi^0}$ (bottom) with $p_{\Sigma^0\pi^0}$ for $K^-$ absorptions in the wall (by $\theta_{\Sigma^0\pi^0}$ we mean the angle between $\textbf{p}_{\Sigma^0}$ and $\textbf{p}_{\pi^0}$ in the laboratory reference frame). The LM component is correlated with masses above $m_{lim}$ (around 1430 $MeV/c^2$ i.e. few $MeV$ below the mass threshold in flight)  and greater angles, the HM component is correlated with masses below $m_{lim}$ and smaller angles. Similar correlations  appear studying $p_{\Sigma^0\pi^0}$ and $m_{\Sigma^0\pi^0}$ as a function of the pion kinetic energy $T_{\pi^0}$. In particular from Fig. \ref{mtvsp_s0pi0} right the LM $p_{\Sigma^0\pi^0}$ component results to be associated with an enhancement at $T_{\pi^0} \sim 70-80 MeV$ (corresponding to higher $m_{\Sigma^0\pi^0}$ values) while the HM $p_{\Sigma^0\pi^0}$ component is associated with $T_{\pi^0} \sim 50-60 MeV$. The pion kinetic energy resolution is $\sigma_{T_{\pi^0}} \sim 12MeV$.


Analogous considerations are also valid for $K^-$ absorption in gas events.

\section{Comparison of the $p_{\Sigma^0\pi^0}$ and $T_{\pi^0}$ distributions with emulsion experiments}\label{Comparing}

The kinematics of the selected events, originating the characteristic structure of the observed distributions, was clarified by comparing our results  with the corresponding distributions reported in \cite{ref:rif1} for $\Sigma^\pm\pi^\mp$ production following $K^-$ absorption at rest in emulsion. In \cite{ref:rif1} the $p_{\Sigma^\pm\pi^\mp}$ distribution (for a compilation of various experiments studying K$^-$ absorption \emph{at rest} in emulsion, with similar angular cuts for the final selected recoil events) is found to be peaked around $200 MeV/c$ (in good agreement with Adair's calculation for the $K^- \, C \rightarrow \Sigma \pi \,\, {}^{11}B$ reaction at rest) analogously to the red curve in Fig. \ref{m_s0pi0_wall}. The red and green curves in Fig. \ref{m_s0pi0_wall} are obtained by splitting the whole sample for $m_{\Sigma^0\pi^0}$ values less or greater than $m_{lim}$ respectively, which clearly separates the LM and HM components of the spectrum.

Moreover, masses below $m_{lim}$ are characterized by $\pi^0$ kinetic energies distributed around $(50-60) MeV$ (see Fig. \ref{mtvsp_s0pi0}) according with the $T_{\pi^\pm}$ distribution observed in \cite{ref:rif1} for K$^-$ absorption at rest (the $T_{\pi^0}$ threshold for $K^-$ absorption at rest in ${}^{12}C$ is $77 MeV$). Higher $T_{\pi^0}$ values are associated to  $m_{\Sigma^0\pi^0} > m_{lim}$ events.

The LM component then results to be associated to in flight absorptions, the momentum of the kaon ($\sim 120 MeV/c$) enabling to access masses above the $m_{lim}$ threshold.

Analogous conclusions were achieved for $K^-$ absorptions in gas events by comparing our spectra with Helium bubble chamber experiments \cite{ref:katz,ref:brun}. 

The red and green spectra in Fig.s \ref{m_s0pi0_wall} and \ref{mtvsp_s0pi0} are also well consistent with MC simulations of at rest and in flight $K^-$ absorption.

\section{Analysis of the background: $\Sigma(1385)$ and internal conversion events}\label{back1}

Two main background sources can affect the final reconstructed events, namely $\Sigma(1385)$ events and internal conversion.
The $\Sigma(1385)$ can not decay into $\Sigma^0\pi^0$ for isospin conservation, but its main decay mode ($\Lambda\pi^0 \, , \, BR=87.0 \pm 1.5 \%$) could constitute a source of background if the photon associated to the $\Sigma^0$ decay by the reconstruction algorithm ($\gamma_3$) was not "genuine". The main cause for neutral cluster misidentification as a photon cluster could be a wrong track to cluster association, nevertheless the time of flight selection ($\chi_t^2$ minimisation) is expected to drastically reduce such occurrence, as confirmed by this analysis.
Internal conversion occurs as a consequence of secondary interactions of the $\Sigma^0$ with nucleons, following the reaction 
$\Sigma^0 \, N \rightarrow \Lambda \, N'$,
which competes with the $\Sigma^0$ decay process in $\Lambda \, \gamma_3$.

Both processes produce the same final state particles $\Lambda \pi^0$. The incidence of background contamination was first tested by a characterization of the time and energy spectra of neutral clusters associated to $\gamma_3$, and a study of the $\Lambda$ momentum distribution. The estimate of the background contribution was obtained by means of dedicated MC simulations of pure signal ($\Lambda(1405) \rightarrow \Sigma^0 \pi^0$) and pure background (of both types) events. Both $K^-$ interactions at rest and in flight were generated in equal number. The various physical processes where weighted with the corresponding branching ratios according to \cite{ref:van-vel} and \cite{ref:katz}.

The percentage of background events entering the final selected sample is then obtained to be  $0.03 \pm 0.01$ for $K^-$ absorptions in the DC wall and $0.03 \pm 0.02$ for $K^-$ absorptions in the DC gas 
As expected the $\Lambda \pi^0$ contamination is small. The $\Sigma^0 \pi^0$ lineshape is then almost free from the $I=1$ $\Sigma(1385)$ contribution, which is one of the  main strengths of this study.

\section{Conclusions}

In this work the absorption of negative kaons on bound protons in the KLOE DC internal wall and the DC gas (mainly ${}^{12}C$ and ${}^4He$ respectively) was investigated, through an exclusive study of the  $\Sigma^0\pi^0$ events produced   in the final state. The analysis was motivated by the need of a deeper understanding of the $\Lambda(1405)$ nature and of its behaviour in nuclear environment. 

The complex experimental condition, involving $K^-$ nuclear absorptions on different nuclear targets both at rest and in flight, was resolved by means of a careful comparative study of the relevant kinematical variables spectra, with the bubble chamber and the emulsion studies of stopped $K^-$ interactions and MC simulations. 
This leads, for the first time, to the identification of in flight contributions to the $m_{\Sigma^0\pi^0}$ spectrum, lying above the kinematical mass limits for processes at rest.

The $\Lambda\pi^0$ background, form direct and internal conversion production mechanisms, was estimated to account for $(3\pm1)\%$ and $(3\pm2)\%$ of the final selected events in the wall and the DC gas respectively. 

This study motivated the realization of a dedicated half-cylinder high purity Carbon (graphite) target in summer 2012. The target was installed inside the KLOE DC, between the beam pipe and the DC entrance wall and an integrated luminosity of $\sim100pb^{-1}$ was collected in the period 6 November - 14 December 2012. Presently the data analysis is undergoing.

\section*{Acknowledgements}


This work is dedicated to Prof. Paul Kienle, who recently left us. He was
constantly helping and supporting me. We will always remember you.

We would like to thank F. Bossi, S. Miscetti, E. De Lucia, A.
Di Domenico, A. De Santis and V. Patera for the guidance in performing
the analyses, and
all the KLOE Collaboration for the fruitful collaboration.

Part of this work was supported
by the European Community-Research Infrastructure Integrating Activity ``Study of Strongly 
Interacting Matter'' (HadronPhysics2, Grant Agreement No. 227431, and HadronPhysics3 (HP3)
Contract No. 283286) under the Seventh Framework Programme of EU.

\end{document}